\documentclass[12pt]{article} %,a4paper
\usepackage{graphicx}
\usepackage{amsmath}
\usepackage{amssymb}
\usepackage[utopia]{mathdesign}
\usepackage{eucal}
 \usepackage{cite}
 \usepackage{hyperref}
\usepackage[height=8.8in,width=6.45in]{geometry}

 \def\inc#1{\includegraphics[scale=.32]{#1}}

\numberwithin{equation}{section}
\newcommand{\cM}{\mathcal{M}}
\newcommand{\cN}{\mathcal{N}}
\newcommand{\cO}{\mathcal{O}}

\newcommand{\bC}{\mathbb{C}}
\newcommand{\bR}{\mathbb{R}}
\newcommand{\bZ}{\mathbb{Z}}

\def\U{\mathrm{U}}
\def\SU{\mathrm{SU}}
\def\fsl{\mathfrak{sl}}
\def\USp{\mathrm{USp}}
\def\SO{\mathrm{SO}}
\def\O{\mathrm{O}}
\def\SU{\mathrm{SU}}

\def\tr{\mathrm{tr}}
\def\diag{\mathop{\mathrm{diag}}}
\def\rank{\mathop{\mathrm{rank}}}
\def\hkq{\mathop{/\!/\!/}}

\begin{document}

\begin{titlepage}

\begin{flushright}
IPMU-14-0029\\
UT-14-5\\
\end{flushright}

\vskip 3cm

\begin{center}
{\Large \bfseries
Moduli spaces of
$\SO(8)$ instantons on smooth ALE spaces  \\[1em]
as Higgs branches of 4d $\cN{=}2$ supersymmetric theories
}

\vskip 2.0cm
  Yuji Tachikawa
\vskip 1.0cm

\begin{tabular}{ll}
  & Department of Physics, Faculty of Science, \\
& University of Tokyo,  Bunkyo-ku, Tokyo 133-0022, Japan, and\\
  & Institute for the Physics and Mathematics of the Universe, \\
& University of Tokyo,  Kashiwa, Chiba 277-8583, Japan\\
\end{tabular}

\vskip 1cm

\textbf{Abstract}

\end{center}

The worldvolume theory of D3-branes probing four D7-branes and an O7-plane on  $\bC^2/\bZ_2$ is given by a supersymmetric $\USp\times\USp$ gauge theory.  
We demonstrate that, at least for a particular  choice of the holonomy at infinity, we can go to  a  dual description of the gauge theory, where we can add a Fayet-Iliopoulos term  describing the blowing-up of the orbifold to the smooth ALE space.  
This allows us to express the moduli space of $\SO(8)$ instantons on the smooth ALE space as a hyperk\"ahler quotient of a flat space times the Higgs branch of a class S theory.  We also discuss a generalization to $\bC^2/\bZ_{2n}$, and speculate how to extend the analysis to bigger SO groups and to ALE spaces of other types.

\medskip
\noindent

\end{titlepage}

%\setcounter{tocdepth}{2}
%\tableofcontents

%%%%%%%%%%%%%%%%%%%%%%%%%%%%%%%%%%%%%%%%%%%%%

\section{Introduction}

The objective of this paper is to revisit the problem of the  gauge theory description of  D3-branes probing a D7-O7 system on smooth asymptotically-locally-Euclidean (ALE) spaces. Let us first recall what the difficulty was. 

\begin{figure}[h]
\[
\inc{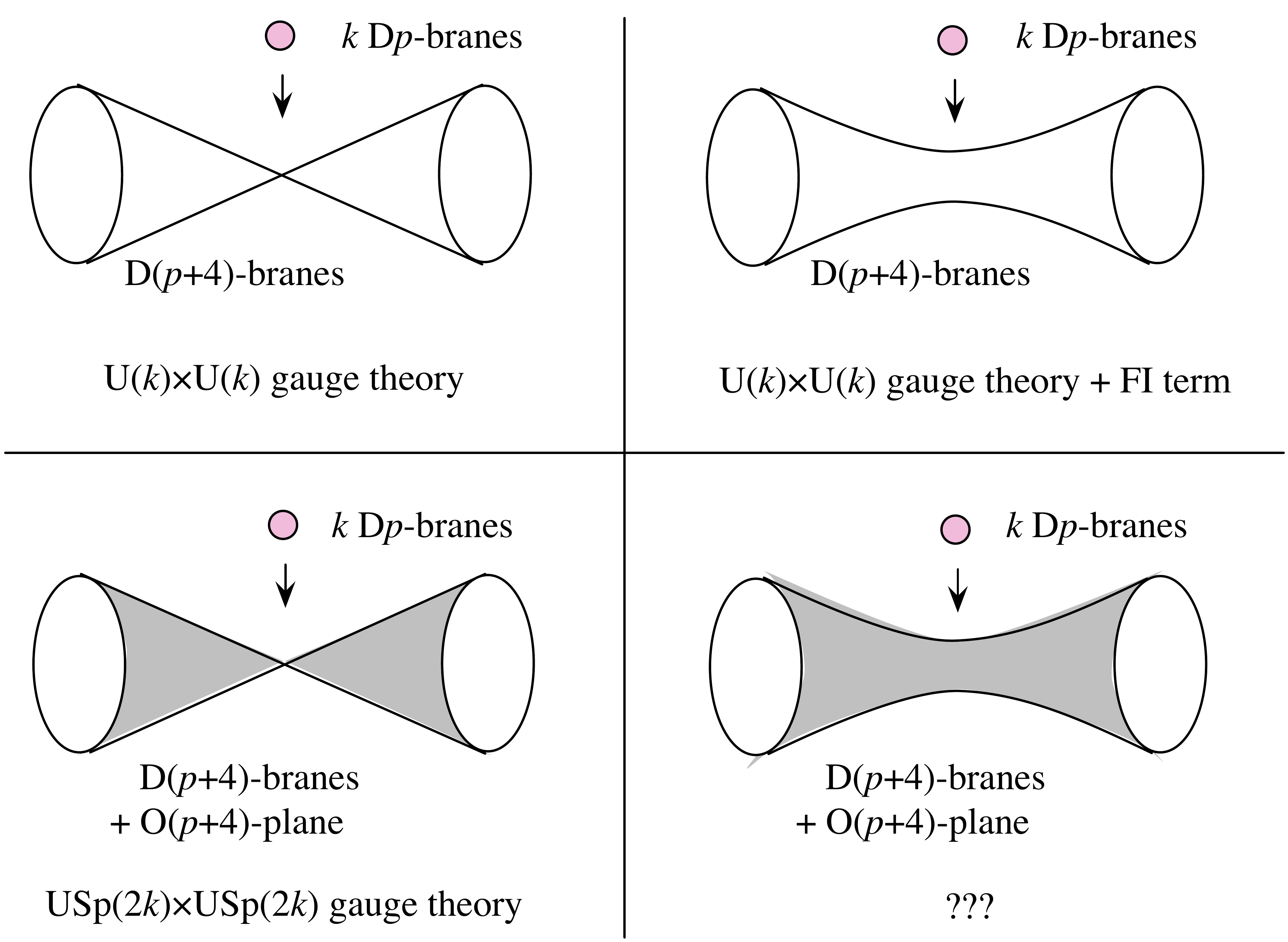}
\]
\caption{D$p$-branes probing D$(p+4)$-branes on $\bC^2/\bZ_2$, with and without an orientifold\label{question}}
\end{figure}

It is by now well-known %\cite{Witten:1995gx} 
that the open-string description of $k$ D$p$-branes probing flat D$(p+4)$-branes, with or without O$(p+4)$-plane, realizes the ADHM construction of the moduli space of instantons. 
When the $(p+4)$-branes are put on an orbifold $\bC^2/\bZ_2$, the world-volume theory of D$p$-branes becomes a quiver gauge theory. Without any O$(p+4)$-plane in place, the gauge group is of the form $\U(k)\times\U(k')$, and the blow-up parameters of the orbifold are given by the FI terms of the theory. This reproduces Kronheimer-Nakajima construction \cite{KronheimerNakajima,Bianchi:1996zj} of unitary instantons on the ALE space $\widetilde{\bC^2/\bZ_2}$, as first shown in \cite{Douglas:1996sw}.

There is a problem with an O$(p+4)$-plane, however.  The gauge group of the system, which can be found by quantizing open strings on the orbifold, is now $\USp(2k)\times \USp(2k')$, for which we cannot add any FI terms.  Still, it is clear geometrically that we can still blow up the orbifold.  We would like to  understand this process better in the gauge theory language and to find a way to describe the moduli space of orthogonal instantons on smooth ALE spaces. 

% wrote up to this point on Jan 19 in the morning.
% started working again at night.

When $p=2$, the gauge theory is three dimensional with $\cN=4$ supersymmetry.  FI terms cannot be added directly to the $\USp\times\USp$ gauge theory, but we can use the 3d mirror description (see e.g.~\cite{Dey:2011pt,Dey:2013nf,Dey:2014tka} for  recent discussions), where the blow-up parameters are visible as hypermultiplet mass terms. The moduli space of orthogonal instantons on a smooth ALE space is then given as the quantum-corrected Coulomb branch of this mirror theory.  In general, describing the Coulomb branch of 3d $\cN=4$ theories is a difficult problem, and therefore this construction does not yet tell us much about the moduli space of orthogonal instantons on the smooth ALE spaces.

When $p=5$, the gauge theory is six dimensional with $\cN=(1,0)$ supersymmetry. Here, as noticed first in \cite{Intriligator:1997kq}, the blow-up parameters are a part of hypermultiplets involved in the transition between the tensor branch and the Higgs branch, about which not much is understood yet either. 

In this paper, we take $p=3$, so that the gauge theory is four dimensional with $\cN=2$ supersymmetry. In the last few years, a significant progress has been made in the understanding of the duality of such systems. We will see that, at least for $\SO(8)$ instantons with a particular holonomy at infinity, we can go to a  dual description of the original gauge theory, where we can add appropriate FI terms. This method will give a description of the moduli space of such instantons as a hyperk\"ahler quotient of a flat space times the Higgs branch of a particular class S theory.  When the instanton number is sufficiently small, the moduli space reduces to a hyperk\"ahler quotient of a flat space times a nilpotent orbit.

In the next section, we study $\SO(8)$ instantons on the blown-up ALE space $\widetilde{\bC^2/\bZ_{2}}$.  
We provide two complimentary approaches leading to the same conclusion: one uses the embedding to string theory directly, and the other uses a field-theoretical infrared duality recently discussed in \cite{Gaiotto:2010jf,Giacomelli:2012ea}.
In Sec.~\ref{sec:2n}, a generalization to $\widetilde{\bC^2/\bZ_{2n}}$ will be described.
We conclude with a discussion in Sec.~\ref{sec:conclusions}, where we speculate how we can extend the analysis to larger $\SO$ groups and to ALE spaces of other types.

\section{$\SO(8)$ instantons on  $\widetilde{\bC^2/\bZ_2}$}\label{sec:2}
\subsection{Basic mathematical facts}
Let us first recall the quiver description of the moduli space of $\SO(N)$ instantons on the orbifold $\bC^2/\bZ_2$, with the holonomy at infinity given by \begin{equation}
\diag(\underbrace{++\cdots+}_\text{$N_+$ times}
\underbrace{--\cdots-}_\text{$N_-$ times})\label{holinf}
\end{equation} where $N=N_++N_-$.  Note that $N_-$ is even. \footnote{A readable account of the moduli space of $\SU(N)$ instantons on the orbifold $\bC^2/\Gamma$ can be found in \cite{Bianchi:1996zj,Fucito:2004ry}. }

The moduli space is given by the Higgs branch of a $\USp(2k_+)\times \USp(2k_-)$ gauge theory, with $N_+$, $N_-$ fundamental half-hypermultiplets for the first and the second gauge factors, and a bifundamental hypermultiplet of the two $\USp$ factors.\footnote{Our convention is that $\USp(2)=\SU(2)$.}  
The holonomy at the origin can be computed  by a method explained e.g.~in Appendix B of \cite{Witten:2009xu}, and is given by \begin{equation}
\diag(\underbrace{++\cdots+}_\text{$N_+'$ times}
\underbrace{--\cdots-}_\text{$N_-'$ times}), 
\end{equation}
where
\begin{equation}
N_+'=N_+-4(k_+-k_-), \quad
N_-'=N_--4(k_--k_+).
\end{equation}

Let us next consider  the orthogonal instantons on the blown-up ALE space $\widetilde{\bC^2/\bZ_2}$, with the  holonomy at infinity given by \eqref{holinf}. The integral of $\tr F\wedge F$ in an appropriate normalization    is  given by \begin{equation}
K=k+\frac{N_-}8
\end{equation} where $k$ is an integer, and the dimension\footnote{We always refer to quaternionic dimensions in this paper.} of the moduli space is \begin{equation}
(N-2)K-\frac{N_+N_-}8
\end{equation}  where the second term is the contribution from the $\eta$ invariant at the asymptotic boundary.  
The second Stiefel-Whitney class of the bundle is determined by the holonomy at infinity, and therefore does not give additional topological data. For more explanations of the facts in this paragraph, see e.g.~Sec.~4 of \cite{Berkooz:1996iz}.

When $N_-$ is a multiple of four, we see that the dimensions of the moduli spaces on the orbifold and the smooth ALE space agree when we take $k_+=k$, $k_-=k+N_-/4$. For unitary instantons on the ALE space, the difference $k_+-k_-$ controls the first Chern class of the bundle, given by $2(k_+-k_-)+N_-$, on the smooth space \cite{KronheimerNakajima}. For orthogonal instantons, on the contrary, the difference $k_+-k_-$ does not correspond to any data of the gauge configuration on the smooth ALE space. 

\subsection{Analysis using string dualities}\label{string}
\subsubsection{T-duality to the D4-D6-O6 system}
We are going to study this system using D3-branes probing $N$ D7-branes and an O7-plane on the orbifold or the smooth ALE space.  For a techincal reason, we choose $N_{\pm}$ and $k_\pm$ so that the resulting four-dimensional  supersymmetric gauge theory is conformal or slightly asymptotically free. This is so that we can apply the field theoretical dualities found in the last few years, starting in \cite{Gaiotto:2009we}. This choice also facilitates the analysis using branes, since the bending of NS5-branes will be (almost) absent.  Concretely, we choose $N_+=N_-=4$, and set
\begin{equation}
(k_+,k_-)=(k,k),\quad \text{or}\quad (k_+,k_-)=(k,k+1) \label{choice}
\end{equation}
 Note that $N_++N_-=8$ corresponds to the familiar choice where the dilation tadpole of the O7-plane is canceled by that of the D7-branes.

\begin{figure}[h]
\[
\inc{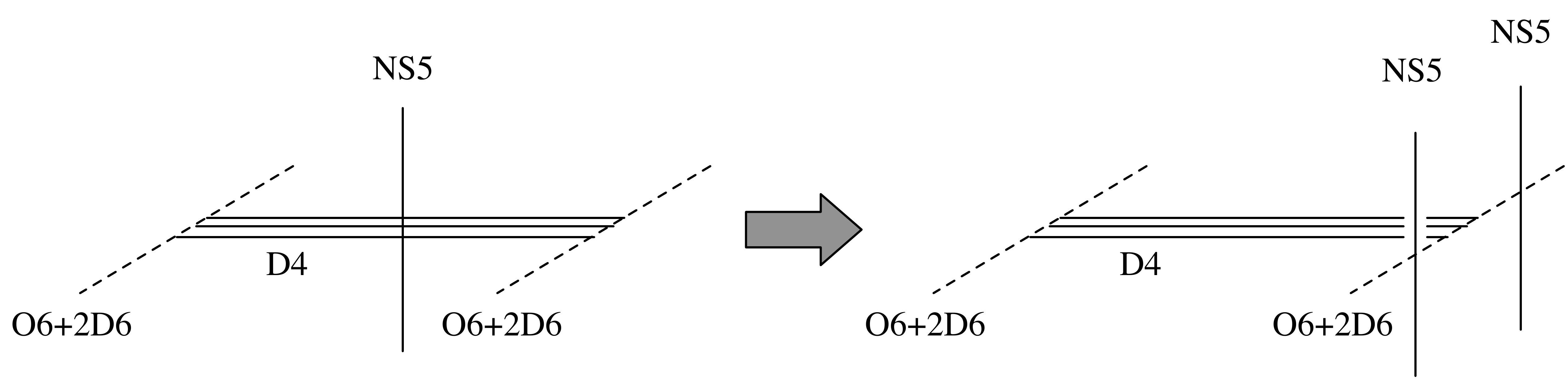}
\]
\caption{Type IIA configuration after a T-duality\label{fig:IIA}}
\end{figure}

We first deform the ALE space to a two-centered Taub-NUT space, around whose $S^1$ fibers we perform the T-duality.  The resulting configuration is shown in Fig.~\ref{fig:IIA}.
The spacetime is of the form $\bR^{3,1}\times (\bC\times S^1)/\bZ_2 \times \bR^3$,
where $\bZ_2$ is the orientifolding action. Every brane fills $\bR^{3,1}$. In addition, the NS5-branes, the D4-branes, and the D6-branes extend along $\bC$, $S^1$, and $\bR^3$, respectively. 
Each of the two O6-planes has four D6-branes on top, corresponding to the choice $N_+=N_-=4$.

Recall that the relative distance along $\bR^3$ between the NS5-brane and its orientifold image is the blow-up parameter of the ALE space. This can be nonzero only when the NS5-brane is on top of the O6-plane, where it meets its mirror image under the orientifolding action, as shown in the same figure.  In terms of the gauge theory describing the dynamics of the D4-branes, this means that the blow-up parameter can only be introduced when one of the gauge couplings is extremely strong. 

\subsubsection{Re-interpretation as a class S construction}

\begin{figure}[h]
\[
\inc{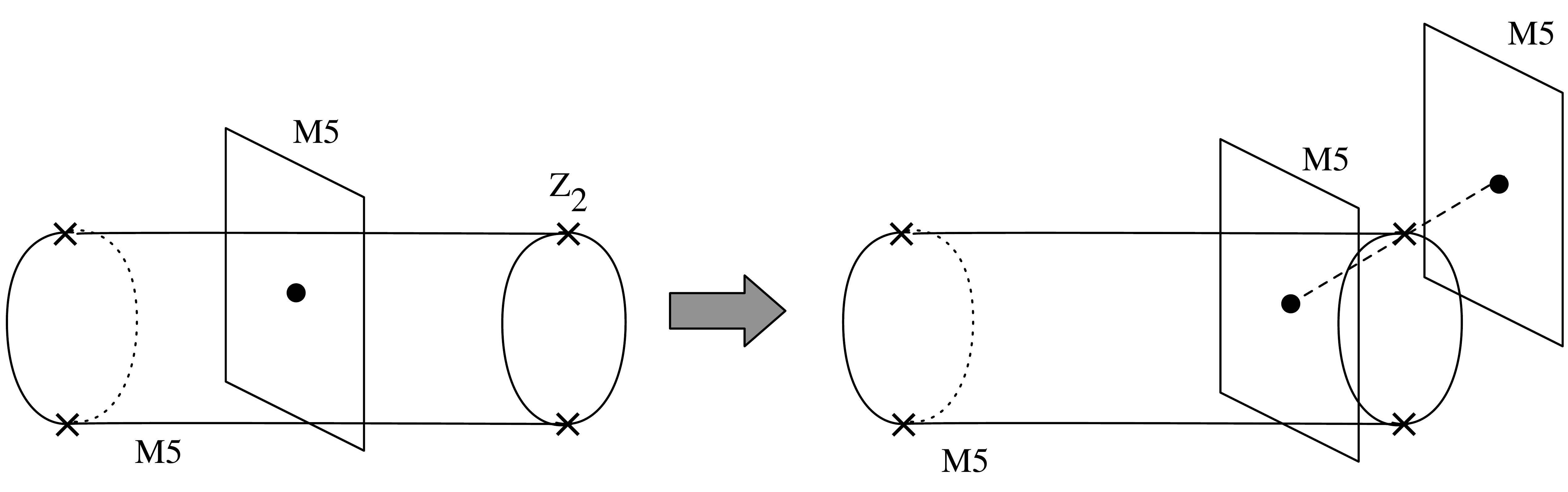}
\]
\caption{Lifted M-theory configuration\label{fig:M}}
\end{figure}

Now, let us lift the set-up to M-theory. When $k_+=k_-=k$, this can be done very easily, since there is no bending of the NS5-brane. The result is shown in Fig.~\ref{fig:M}.
The spacetime is now of the form $\bR^{3,1}\times (\bC\times T^2)/\bZ_2 \times \bR^3$; two O6-planes become four $\bZ_2$ singularities.  When the vertical M5-brane (the one not wrapping the M-theory circle)  is on top of a $\bZ_2$ singularity,  
we can separate it into two, and each piece can be moved along $\bR^3$ independently. Their relative distance is the blow-up parameter. 

\begin{figure}[h]
\[
\inc{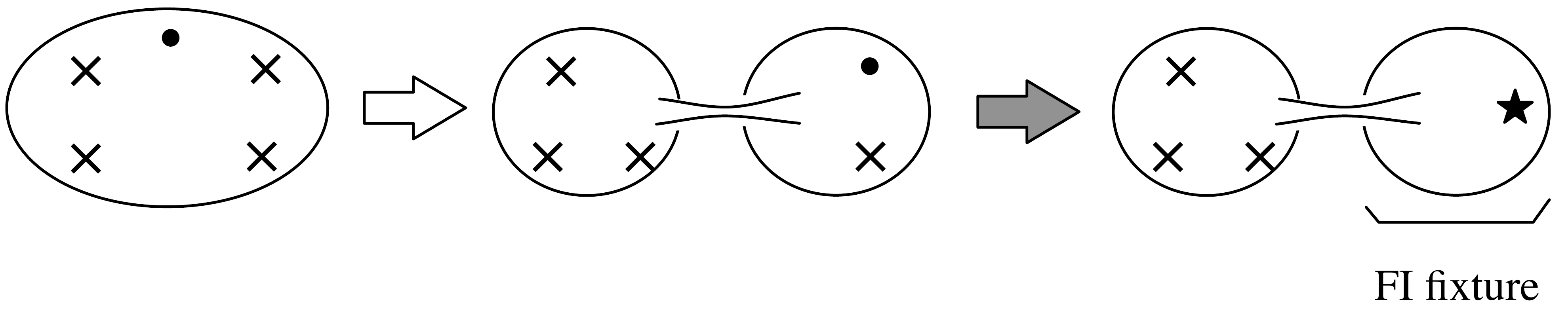}
\]
\caption{The change in the ultraviolet curve under the process. The symbols $\times$, $\bullet$ are for the puncture of type $[k^2]$ and for the simple puncture, respectively. The symbol $\star$ is for a new type of puncture introducing FI-like deformation.  \label{fig:gaiotto}}
\end{figure}

Let us study this process from the point of view of the class S-theory. We have $2k$ M5-branes wrapping $S^2\simeq T^2/\bZ_2$, intersected by a vertical M5-brane. Using the standard rules \cite{Gaiotto:2009we,Nanopoulos:2009xe,Chacaltana:2010ks}, we know that a $\bZ_2$ singularity is a puncture of type $[k^2]$ and an intersection with a vertical M5-brane is a simple puncture. 
Then, when the vertical M5-brane comes very close to the $\bZ_2$ singularity, we can go to a dual frame, as shown by a white arrow in Fig.~\ref{fig:gaiotto}.

We now have a weakly-coupled dual $\SU(2)$ gauge group from a long neck.
The three-punctured sphere on the right hand side in the figure, containing a simple puncture and a puncture of type $[k^2]$ corresponds to an empty matter content.\footnote{This is true only when $k>1$. When $k=1$ a slight modification of the analysis is necessary, since the three-punctured sphere on the right hand side also gives a trifundamental.}
Putting the vertical M5-brane on top of the $\bZ_2$ singularity is to make the coupling of the dual $\SU(2)$  gauge group to be exactly zero. This is not a continuous process in the field theory language. We  therefore represent the process of separating the M5-brane and its mirror image on the $\bZ_2$ singularity  by gluing in a different sphere on the right hand side, with a new puncture representing the separated M5-branes on the $\bZ_2$ singularity. 
We showed this procedure by a black arrow in Fig.~\ref{fig:gaiotto}. 
Let us call the new contribution  on the right hand side of the neck as an FI fixture.

\subsubsection{Identification of the new contribution}
\begin{figure}[h]
\[
\inc{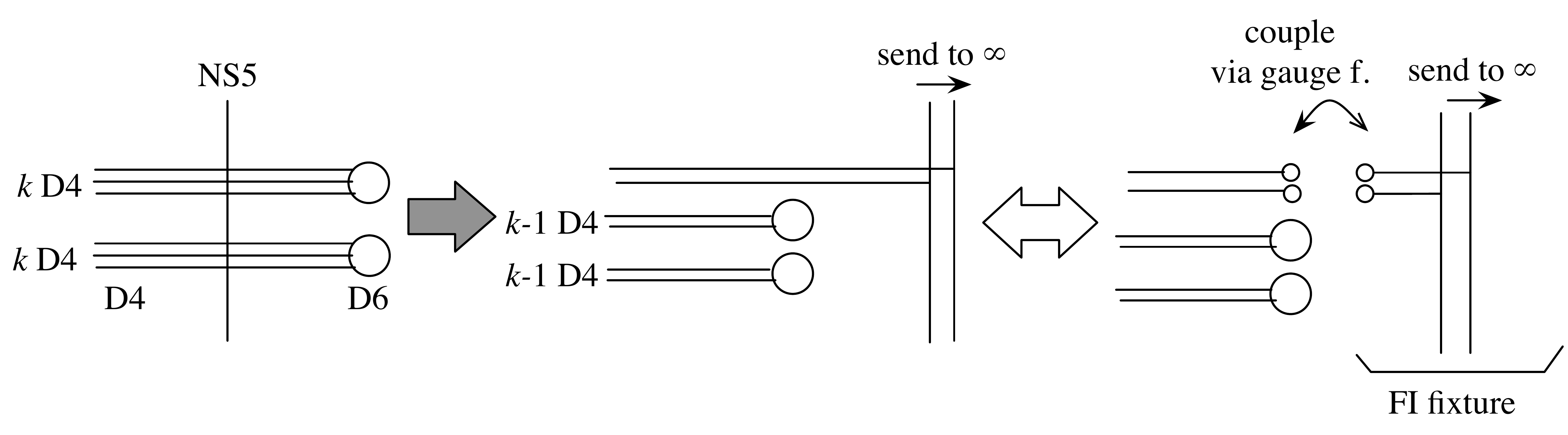}
\]
\caption{Behavior close to the M-theory $\bZ_2$ singularity, after reduction to Type IIA\label{fig:reduction}}
\end{figure}

 To understand what the FI fixture does, we can reduce the system close to the $\bZ_2$ singularity to Type IIA theory with a different choice of M-theory circle, such that the $\bZ_2$ singularity sits at the tip of the cigar. The result is shown in Fig.~\ref{fig:reduction}.  When the vertical M5 is not exactly on top of the singularity, the outcome of the reduction is an NS5-brane, intersecting $2k$ D4-branes ending on two D6-branes in equal numbers. When the vertical M5 is exactly on top of the singularity and separated along $\bR^3$,  the outcome of the reduction is essentially given by the Hanany-Witten effect: Now $k-1$ D4-branes end on each D6-brane, and  there are in addition two semi-infinite D4-branes whose boundary condition is given by the separation along $\bR^3$. To visualize the very-weak $\SU(2)$ group, we can artificially cut the two D4-branes by introducing four D6-branes, remembering that we need to couple the two resulting $\SU(2)$ flavor symmetries by a gauge symmetry. With this process, we clearly see the brane realization of the FI fixture. 

Luckily, this brane set-up realizing the FI fixture was already studied in \cite{Gaiotto:2008ak}.
Field theoretically, it is given by a $\U(1)$ gauge theory with two flavors with a FI term,  and its Higgs branch is just $\widetilde{\bC^2/\bZ_2}$.  Coming back to Fig.~\ref{fig:gaiotto}, we replaced the empty matter content on the right hand side with a one-dimensional Higgs branch. Therefore, the process shown by the black arrow there adds one  dimension to the Higgs branch. 

\begin{figure}[h]
\[
\inc{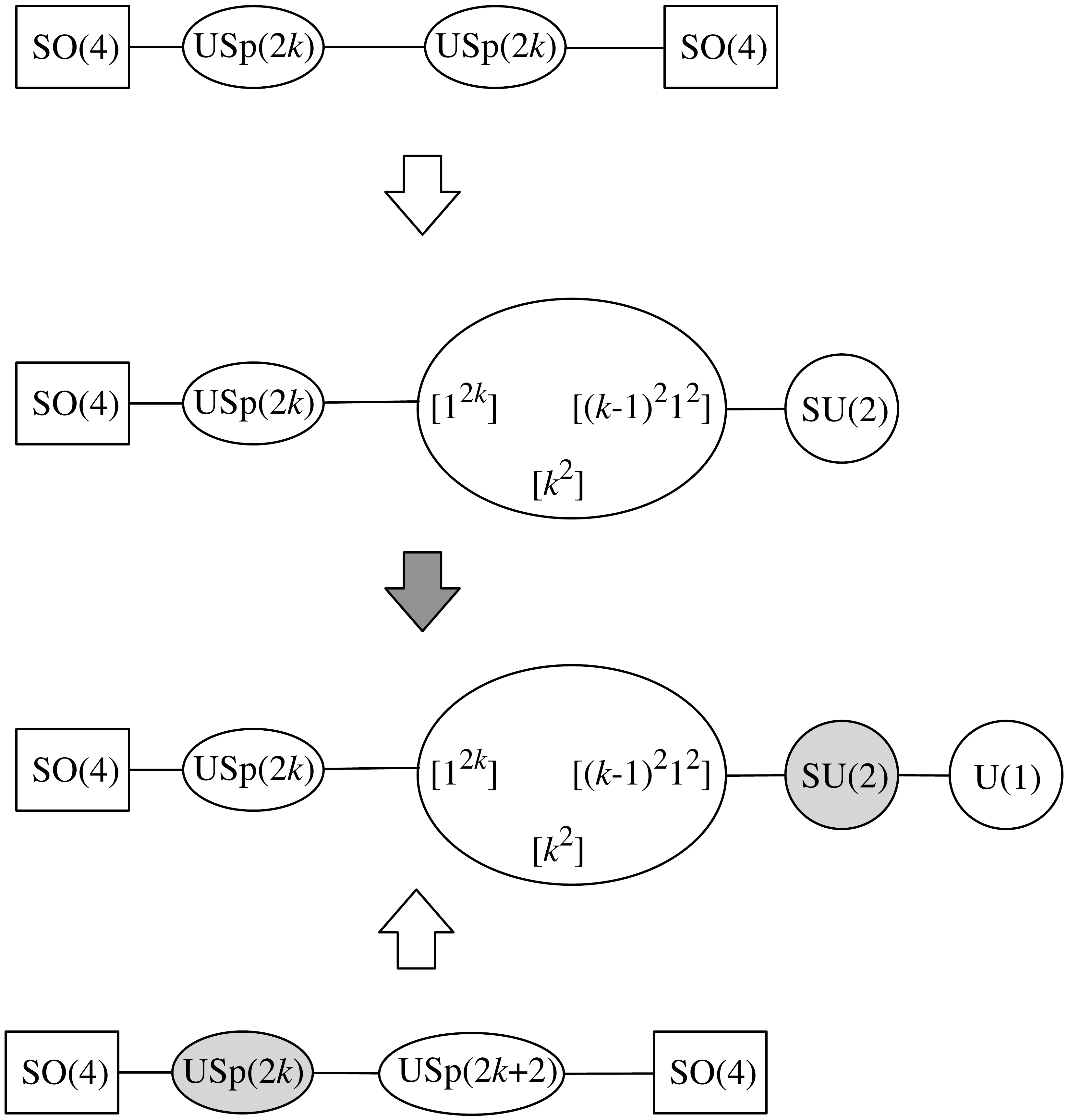}
\]
\caption{Summary of the gauge-theoretic operations. Infrared-free gauge groups are shaded.\label{fig:summary}}
\end{figure}

\subsection{Summary of the procedure}

Let us summarize the process described so far in a field theoretical language, see the first three rows of Fig.~\ref{fig:summary}.  \begin{itemize}
\item We start from a $\USp(2k)\times \USp(2k)$ gauge theory, with two fundamental hypermultiplets for each $\USp$ group and a bifundamental hypermultiplet. The dimension of the Higgs branch is $6k$.  We want to make the right $\USp(2k)$ very strongly coupled.
\item We go to an S-dual frame on the right $\USp(2k)$ gauge group: this involves a three-punctured sphere with a full puncture, a puncture of type $[k^2]$, and another puncture of type $[(k-1)^21^2]$. The last puncture has a flavor symmetry $\SU(2)$ associated to the parts $1^2$ of the last puncture, to which the $\SU(2)$ gauge multiplet couples weakly. 
\item We take  a $\U(1)$ gauge theory with two flavors with a FI term $\xi$, and couple its $\SU(2)$ flavor symmetry to the $\SU(2)$ gauge multiplet we already have. Note that the $\SU(2)$ gauge multiplet is now infrared free. 
The dimension of the Higgs branch is $6k+1$. 
This is the correct dimension of the moduli space of $\SO(8)$ instantons with the holonomy at infinity $\diag(+,+,+,+,-,-,-,-)$. 
\end{itemize}

The Higgs branch $\tilde{\cM}_{k,\xi}$ of this final system, from our chain of string dualities, should give
  (a component of) the moduli space of $\SO(8)$ instantons on the smooth ALE space $\widetilde{\bC^2/\bZ_2}$ with the prescribed holonomy at infinity. 
 Here $k$ is the instanton number and $\xi$ is the blow-up parameter of the smooth ALE space.
  
 For general $k$, it is a hyperk\"ahler quotient of a flat space times the Higgs branch of the class S theory on a three-punctured sphere. This is unfortunately not very explicit yet.\footnote{Ginzburg and Kazhdan have an unpublished manuscript in which the Higgs branch of these theories are constructed as holomorphic symplectic varieties \cite{GinzburgKazhdan}.}

For $k=1$ and $k=2$, the construction becomes completely explicit. 
For $k=1$, we need to make a small modification as was mentioned in the last footnote: 
the rightmost $\SU(2)$ is coupled to another three-punctured sphere.  Then the theory is 
an $\SU(2)\times \SU(2)\times \U(1)$ gauge theory, 
with bifundamental hypermultiplets for consecutive gauge groups,
and additional two flavors for each $\SU(2)$ factor. 
The Higgs branch is a hyperk\"ahler quotient of a vector space.
For $k=2$, the Higgs branch of the three-puncture sphere with punctures $[1^4]$, $[1^4]$ and $[2^2]$ is the minimal nilpotent orbit of $E_7$. Then the Higgs branch of the total system is the hyperk\"ahler quotient of the minimal nilpotent orbit of $E_7$ times a vector space by $\USp(4)\times \SU(2)\times\U(1)$. In both cases, the blow-up parameter of the ALE space is given by the value of the moment map for $\U(1)$.

\subsection{A more field-theoretical approach}\label{another}

This final gauge theory we arrived at after a lengthy analysis using string theory dualities can also be directly obtained field-theoretically, starting from the choice $(k_+,k_-)=(k,k+1)$ in \eqref{choice}. The  gauge theory which we use as the new starting point is shown in the fourth row of Fig.~\ref{fig:summary}. Note that the $\USp(2k)$ gauge multiplet is infrared free with $2k+4$ flavors, while $\USp(2k+2)$ gauge multiplet is asymptotically free with $2k+2$ flavors. 

\subsubsection{The infrared dual of the $\USp(2k+2)$ theory}
Let us first focus on the asymptotically free part. The strongly-coupled dynamics of $\USp(2k+2)$ with $2k+2$ flavors was analyzed in \cite{Giacomelli:2012ea}, as an extension of the work \cite{Gaiotto:2010jf}.  
Here we quote the results of \cite{Giacomelli:2012ea}, with additional comments on the Higgs branch.

The case $k=0$ is the familiar $\SU(2)$ theory with $N_f=2$.
Classically, the Higgs branch has two components, each of which is $\bC^2/\bZ_2$, joined at the origin. 
Quantum mechanically,  we have two singular points on the Coulomb branch, at which a Higgs branch component  emanates. The two components of the Higgs branch, together with the two singular points on the Coulomb branch where they touch, are exchanged under the parity of the flavor symmetry $\O(4)$. 

For general $k$,   the Higgs branch classically is the nilpotent orbit of $\O(4k+4)$ of type $[2^{2k+2}]$.  (For a discussion of the nilpotent orbits of orthogonal groups, see e.g.~\cite{CollingwoodMcGovern}.) This is a very even orbit, and consists of two components, exchanged by the parity of  $\O(4k+4)$.
Quantum mechanically,  on the Coulomb branch, we have two most singular points, at which each of the two components of the Higgs branch touches. 
There, the infrared limit is captured by the following system:
\begin{itemize}
\item First, take a class S theory of type $\SU(2k)$ on a  sphere with three punctures, of type $[1^{2k}]$, $[k^2]$ and $[(k-1)^21^2]$, respectively.  Let us call this the matter sector A. This has  an $\SU(2)$ flavor symmetry associated to the parts $1^2$ of the  last puncture. 
\item Second, take  two free hypermultiplets and couple them to $\U(1)$ gauge multiplet with zero FI term. This has $\SU(2)$ flavor symmetry. Let us call this the matter sector B.
\item We then couple the matter sectors A and B by an $\SU(2)$ gauge multiplet.
\end{itemize} 

At this stage, we realized a very-even nilpotent orbit $[2^{2k+2}]$ of $\SO(4k+4)$ in terms of a hyperk\"ahler quotient by $\SU(2)\times \U(1)$. Note that we can introduce the FI term for the $\U(1)$ gauge multiplet; this should deform the nilpotent orbit to a nearby coadjoint orbit of the same dimension. 

\subsubsection{The infrared dual of the total system}

We can then carry over this result to analyze the strongly-coupled physics of the $\USp(2k)\times \USp(2k+2)$ gauge theory we are interested in, by just adding the $\USp(2k)$ gauge multiplet together with two flavors for it.  The result is, again, given by the theory shown in the third row of Fig.~\ref{fig:summary}, albeit with zero FI term for the $\U(1)$ gauge multiplet. 
We denote the Higgs branch of the theory on the third row by $\tilde\cM_{k,\xi}$, where $\xi$ is the FI parameter.
The discussion above shows that the Higgs branch of the $\USp(2k)\times \USp(2k+2)$ theory, i.e.~the moduli space of $\SO(8)$ instantons on $\bC^2/\bZ_2$ with a holonomy at infinity $\diag(+,+,+,+,-,-,-,-)$ and with a trivial holonomy at the origin, is given by two copies of $\cM_{k,0}$.

% wrote up to this point. It's 2:30 am, Jan, 20.

At present, the author does not have a good argument purely within this second approach why the FI term for the $\U(1)$ gauge group in the infrared dual description can be identified with the blow-up parameter of the ALE space.  Once such an explanation is given, the approach here would give a much quicker way to derive the moduli space of $\SO(8)$ instantons on the smooth ALE space. 

\section{$\SO(8)$ instantons on $\widetilde{\bC^2/\bZ_{2n}}$}\label{sec:2n}

In the string-theoretic approach taken in Sec.~\ref{string},
we can  generalize fairly easily   the analysis in the last section to $\SO(8)$ instantons on the smooth ALE space $\widetilde{\bC^2/\bZ_{2n}}$ with the holonomy at infinity $\diag(+,+,+,+,-,-,-,-)$.
The result, when viewed from a field-theoretical approach in Sec.~\ref{another}, gives a slightly new class of infrared dual description of supersymmetric gauge theories.
This section is mainly meant to describe this latter field-theoretical phenomenon. 

\subsection{String-theoretic analysis}
We start by considering the gauge theory with gauge group $\USp(2k)\times \U(2k)^{2n-2} \times \USp(2k)$,
bifundamental matter fields between two consecutive gauge factors, 
and two additional fundamentals for each of $\USp(2k)$ gauge factors. 
This is the gauge theory describing $k$ D3-branes probing four D7-branes and one O7-plane on $\bC^2/\bZ_{2n}$, with the holonomies at infinity and at zero both given by $\diag(+,+,+,+,-,-,-,-)$.

\begin{figure}[h]
\[
\inc{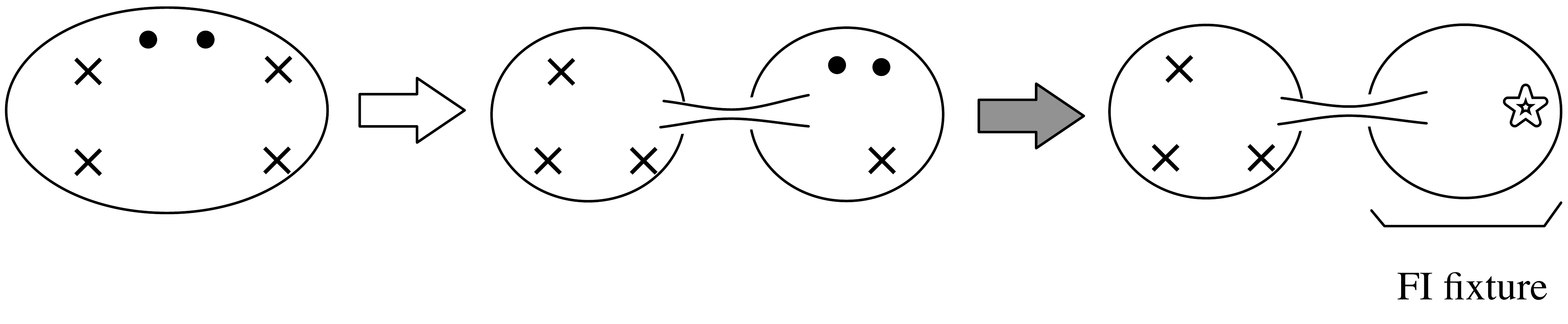}
\]
\caption{The change in the ultraviolet curve. Here we took $n=2$.\label{fig:gaiotton}}
\end{figure}

We follow the same steps as we did in Sec.~2. First, we take the T-dual, and we lift the resulting configuration to M-theory.
The system is described by a class S theory of type $\U(2k)$, put on a sphere with four punctures of type $[k^2]$ and $n$ simple punctures. The process of putting all $n$ vertical M5-branes on top of the $\bZ_2$ singularity can be decomposed into two steps. Namely, we first bring all $n-1$ simple punctures to one puncture of type $[k^2]$, and then replace it with a FI fixture,  see Fig.~\ref{fig:gaiotton}.  We assume $k>n$ for simplicity. 

\begin{figure}[h]
\[
\inc{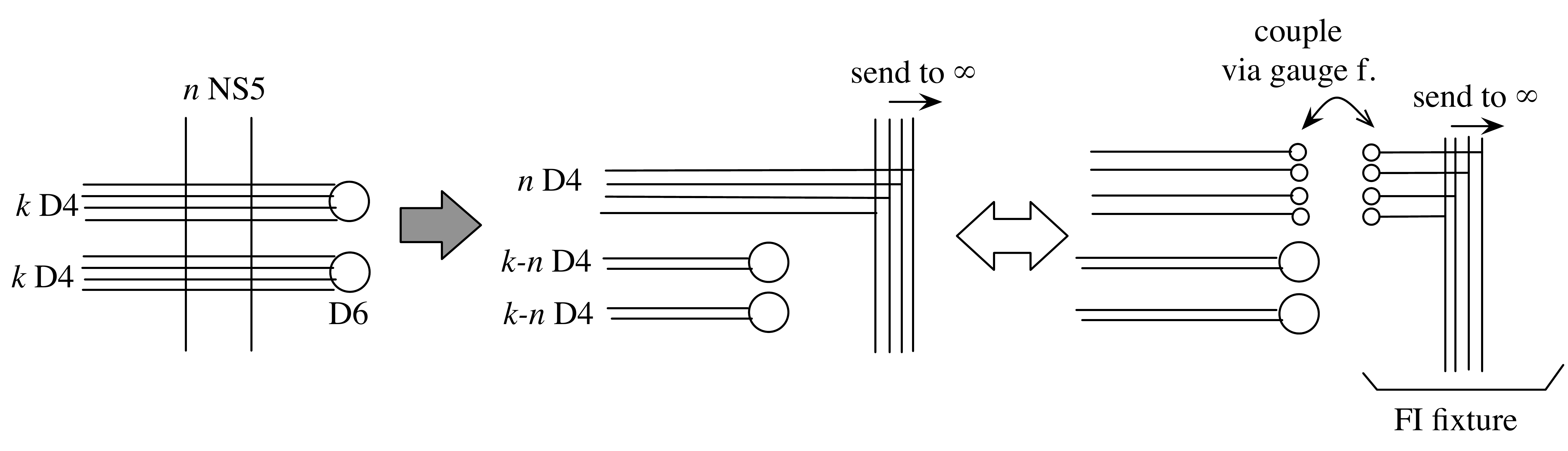}
\]
\caption{Behavior close to the M-theory $\bZ_2$ singularity, after reduction to Type IIA\label{fig:reductionn}}
\end{figure}

The nature of the FI fixture can be found by reducing the system to a Type IIA setup around the $\bZ_2$ singularity, as shown in Fig.~\ref{fig:reductionn}. One finds that the FI fixture can be thought of as a theory with gauge group $\U(2n-1)\times \U(2n-2)\times\cdots \times\U(1)$, with bifundamental hypermultiplets between two consecutive gauge groups, and with additional $2n$ flavors for the first $\U(2n-1)$ group. 
This theory has $2n-1$ Fayet-Iliopoulos parameters $\vec \xi$. When they are all zero, the Higgs branch is the nilpotent orbit $\cN_{A_{2n-1}}$ of $\fsl(2n)$, and when they are turned on, the Higgs branch is  a semisimple orbit $\cO_{A_{2n-1},\vec\xi}$ of $\fsl(2n)$. The $2n-1$ complex FI parameters control the conjugacy class of the orbit, and therefore there is an action of the Weyl group of $\SU(2n)$ on the $2n-1$ FI parameters.  This matches the number of the blow-up parameters for $\bC^2/\bZ_{2n}$, and the action of the Weyl group of $\SU(2n)$ on them.

\begin{figure}[h]
\[
\inc{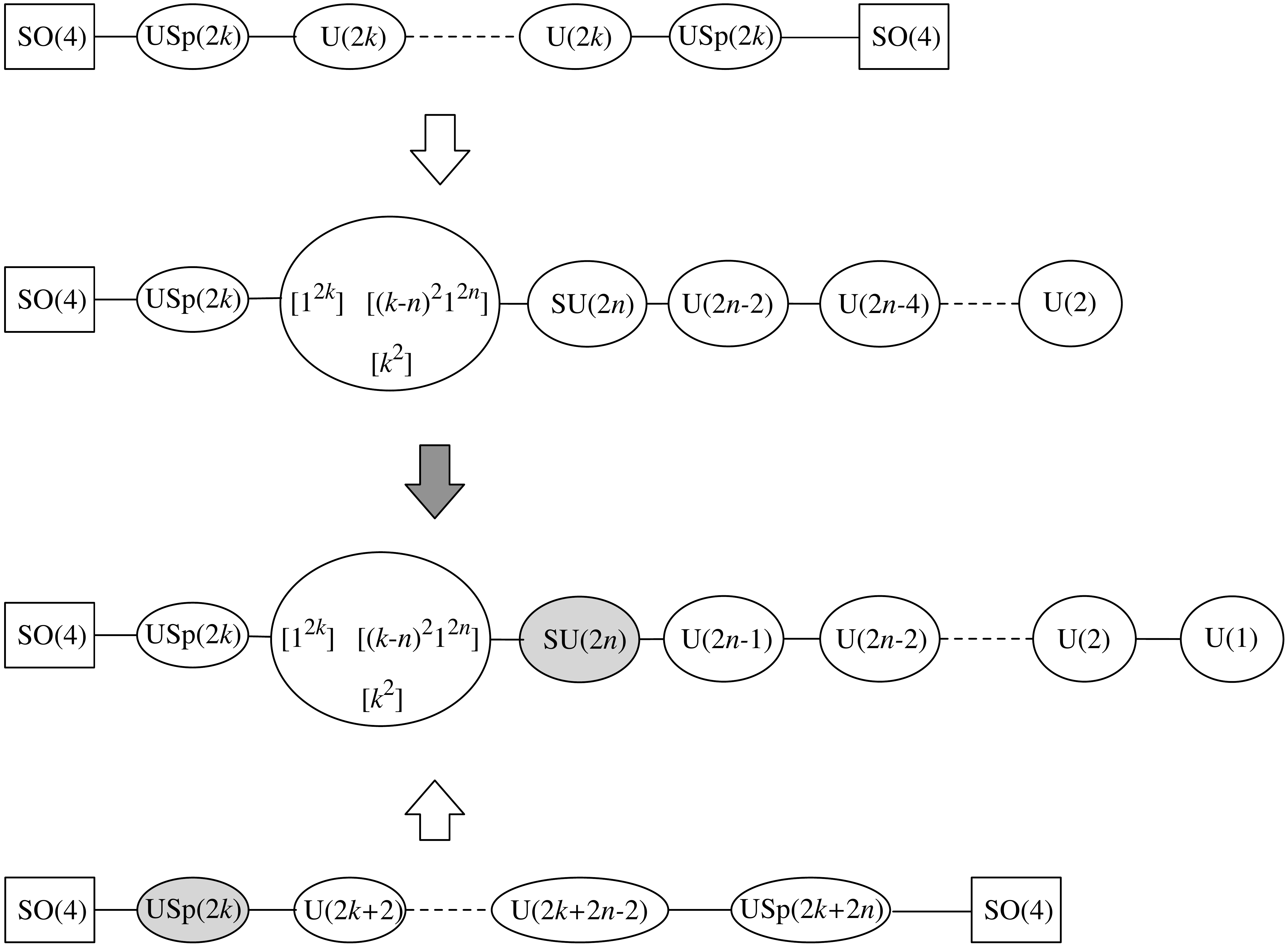}
\]
\caption{Summary of the gauge-theoretic operations\label{fig:summaryn}}
\end{figure}

Let us summarize the process described so far in a field theoretical language, see the first three rows of Fig.~\ref{fig:summaryn}.  \begin{itemize}
\item We start from a $\USp(2k)\times \U(2k)^{n-1}\times \USp(2k)$ gauge theory, with two fundamental hypermultiplets for each $\USp$ group and a bifundamental hypermultiplet for each consecutive gauge groups. The dimension of the Higgs branch is $6k$.  
\item We go to an S-dual frame.  Using the standard techniques of the class S analysis, we find that the resulting theory consists of \begin{itemize}
	\item $\USp(2k)$ group coupled to two fundamentals,
	\item which is coupled further to a class S theory of type $\SU(2k)$ on a sphere with three punctures, of type $[1^{2k}]$, $[k^2]$, and $[(k-n)^21^{2n}]$,
	\item whose $\SU(2n)$ symmetry associated to the parts $1^{2n}$ of the last punctureis coupled  via an $\SU(2n)$ gauge multiplet, 
	\item to another gauge theory with gauge group $\U(2n-2)\times \U(2n-4)\times\cdots\times \U(2)$, with bifundamental hypermultiplets between two consecutive gauge groups, and with additional $2n$ flavors for the first $\U(2n-2)$ group. 
	\end{itemize}
\item We now replace the last item, namely the $\U(2n-2)\times \cdots \times \U(2)$ gauge theory,  
with the gauge theory representing the FI fixture. This is, as explained above, given by a gauge theory with $\U(2n-1)\times \U(2n-2)\times\cdots\times \U(1)$, with bifundamental hypermultiplets between two consecutive gauge groups, and with additional $2n$ flavors for the first  group. 
\end{itemize}

The Higgs branch of the last theory is of dimension $6k+n$, which agrees with the dimension of  the moduli space of $\SO(8)$ instantons with the holonomy at infinity $\diag(+,+,+,+,-,-,-,-)$ on the smooth ALE space $\widetilde{\bC^2/\bZ_{2n}}$.

\subsection{Field-theoretic analysis}
Let us instead consider the quiver gauge theory with gauge group $\USp(2k)\times \U(2k+2)\times \U(2k+4)\times \cdots \times \U(2k+2n-2)\times \USp(2k+2n)$, with bifundamental hypermultiplets between consecutive gauge groups and additional two flavors for each of $\USp$ groups. This is shown in the last row of Fig.~\ref{fig:summaryn}.

The Higgs branch is the moduli space of $\SO(8)$ instantons on $\bC^2/\bZ_{2n}$, with the holonomy at infinity given by $\diag(+,+,+,+,-,-,-,-)$, and a trivial holonomy at the origin. The dimension of the Higgs branch is $6k+n$, which agrees with that of the moduli space of $\SO(8)$ instantons on the smooth ALE space $\widetilde{\bC^2/\bZ_{2n}}$ with the same holonomy at infinity.  Then it is likely that the gauge theory we obtained in the stringy approach will arise as an infrared description close to the most singular points on the Coulomb branch of this fourth theory, as in Sec.~\ref{another}. 

Here we show it is indeed the case.  The coupling of the leftmost $\USp(2k)$ gauge factor is infrared free, so we just neglect them and consider the $\U(2k+2)\times \cdots \U(2k+2n-2) \times \USp(2k+2n)$ gauge theory, with bifundamental hypermultiplets between two consecutive gauge groups and additional $2k$, $2$ flavors for $\U(2k+2)$ and $\USp(2k+2n)$, respectively. 

Its Seiberg-Witten curve is known with all mass parameters turned on \cite{Argyres:2002xc}, but the form is somewhat unwieldy. When the $\SO(4)$ mass parameters are zero, the curve can be embedded into an orbifold of the $(v,t)$ space under the action $(v,t)\to (-v,1/t)$.
The equation of the curve is given by \begin{equation}
P_{n}(v)+ c_{1} (tP_{n-1}(v)+t^{-1}P_{n-1}(-v)) + \cdots + c_n(t^{n} P_0(v)+t^{-n} P_0(-v))=0 \label{curve-org}
\end{equation} with the standard Seiberg-Witten differential $\lambda=vdt/t$. 
Here, $P_j(v)$ is a polynomial of degree $2k+2j$ whose highest coefficient is one, and $P_n(v)=P_n(-v)$. The coefficients of $P_n(v)$ are the Coulomb branch parameters of $\USp(2k+2n)$ gauge multiplet, and those of $P_{j}(v)$ for $j=1,\ldots, n-1$ are the Coulomb branch parameters of $\U(2k+2j)$, and finally those of $P_0(v)$ are the mass parameters for the $\U(2k)$ flavor symmetry. The coefficients $c_1$ to $c_n$ encode the gauge coupling parameters. 

By tuning all the Coulomb branch parameters, we can make the Seiberg-Witten curve singular  at the orbifold fixed point $t=\pm1$. Let us choose $t=1$ for concreteness. 
Say $P_n(v)=v^{2k+2n}+ U v^{2k+2n-2} + \cdots$ where $U$ is the dimension-2 Coulomb branch parameter of $\USp(2k+2n)$ theory. Then, in \eqref{curve-org}, the coefficient of the $v^{2k+2n-2}$ is $c_1(t+t^{-1}) + U$, and the choice $U=-2c_1$ makes the curve more singular. 
We can continue this process, and make the curve very singular there.  

\begin{figure}[h]
\[
\inc{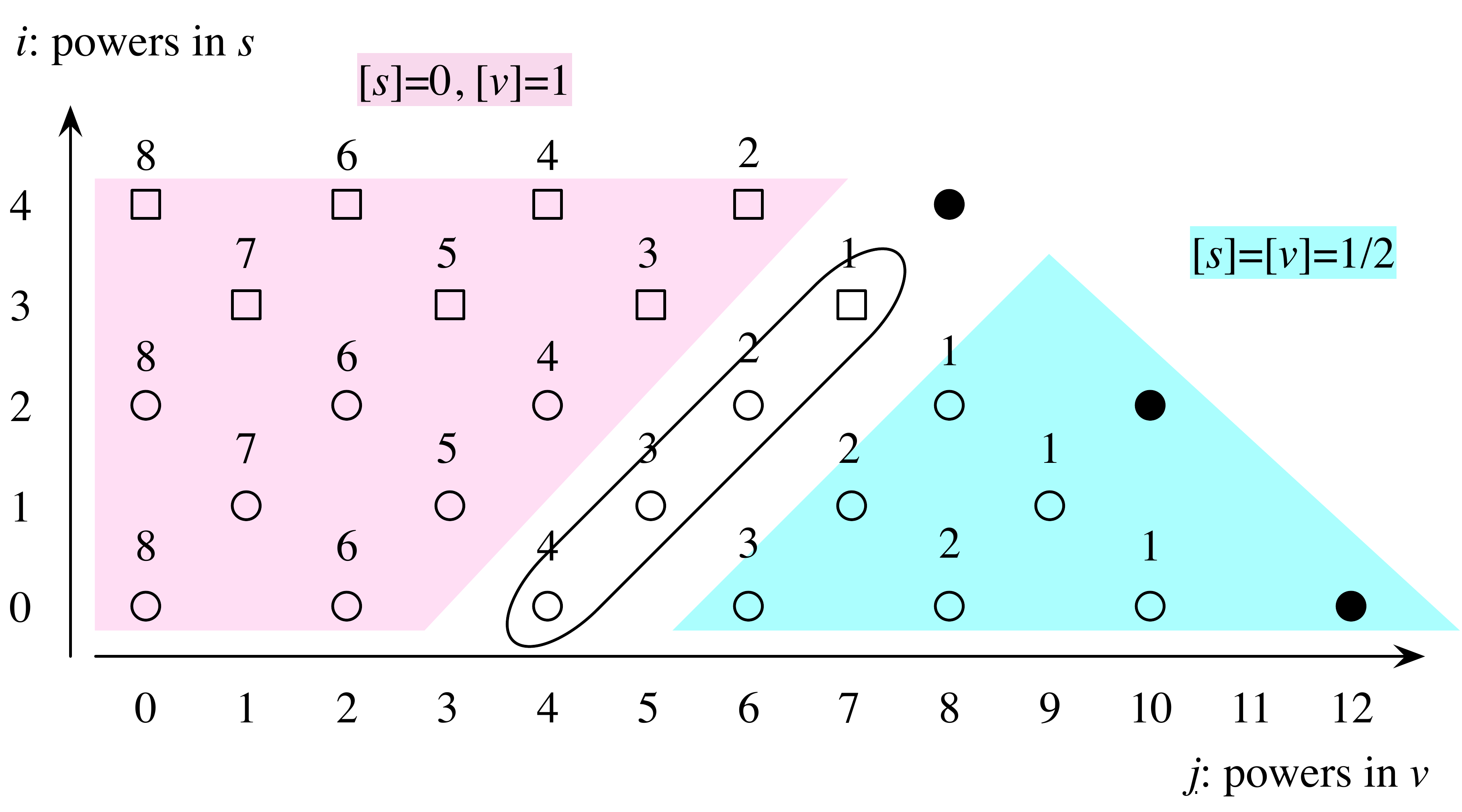}
\]
\caption{Terms in the Seiberg-Witten curves in the singular limit, for $k=4$ and $n=2$. The symbols $\bullet$, $\circ$, $\square$ represent the couplings, the Coulomb branch parameters, and the mass terms, respectively. Powers of $\epsilon$ in \eqref{scaling} is also given at each points.  The three regions shaded or enclosed are for $j-i>2k-2n$, $j-i=2k-2n$, $j-i<2k-2n$.  The symbols $[s]$ and $[v]$ are the scaling dimensions of $s$ and $v$. \label{fig:terms}}
\end{figure}

Expanding $t\sim 1+s$ with very small $s$, the local form of the curve close to $(v,s)=(0,0)$  is given as \begin{equation}
0= v^{2k+2n} + c'_1 s^2v^{2k+2n-2}+ \cdots+ c'_n s^{2n} v^{2k}+ \sum u_{i,j} s^i v^j \label{curve-limit}
\end{equation} where the summation is over the following pairs  $(i,j)$ of non-negative integers: \begin{equation}
i+j\in 2\bZ, \quad i+j < 2k+2n, \qquad i \le  2n.
\end{equation} We perform the identification $(s,v)\simeq (-s,-v)$. 
 The differential is $vds$. 
 In \eqref{curve-limit}, the coefficients $c'_i$ encode (a remnant of) the original gauge couplings. Among $u_{i,j}$, those with $i=2n$ and $i=2n-1$ are the mass parameters for $\U(2k)$, and the rest are the Coulomb branch parameters. 
 They are displayed in Fig.~\ref{fig:terms}, for $k=4$, $n=2$.

Now we look for an appropriate way to scale the parameters, as in \cite{Gaiotto:2010jf,Giacomelli:2012ea}, so as to keep the mass parameters for the non-Abelian flavor symmetry to have canonical dimensions, and to keep as many terms as possible. 
Given a very small number $\epsilon$, a consistent way is to take \begin{equation}
u_{i,j} \sim \begin{cases}
\epsilon^{2k-j} & \text{if $j-i\le 2k-2n$,}\\
\epsilon^{k+n-(i+j)/2} & \text{if $j-i\ge 2k-2n$.}
\end{cases}\label{scaling}
\end{equation} Note that when $j-i=2k-2n$, the two scalings given above both give $u_{2n-l,2k-l}\sim \epsilon^l$.

Then we find three regions on the Seiberg-Witten curve: \begin{itemize}
\item In the region $s\sim 1$, $v\sim \epsilon$, only the terms in \eqref{curve-limit} with $j-i\le 2k-2n$ survive. As the differential is $\sim sdv$, we can assign $\epsilon$ scaling dimension 1.
The same curve arises when we study the strongly-coupled limit of  the superconformal $\USp(2k)\times \U(2k)^{n-1}\times \USp(2k)$ theory we treated earlier. We can thus identify this theory as the class S theory of type $\SU(2k)$, on a sphere with three punctures of type $[1^{2k}]$, $[k^2]$, and $[(k-n)^2 1^{2n}]$.  The parameters $u_{2n-l,2k-l}$ are the mass parameters for the flavor symmetry $\U(2n)$ associated to the parts $1^{2n}$ of  the last puncture. Let us call this the matter sector A.
\item In the region $ s \sim  \epsilon^{1/2}$, $v\sim \epsilon^{1/2}$ , only the terms in \eqref{curve-limit} with $j-i\ge 2k-2n$ survive.  Again, we find $\epsilon$ has scaling dimension 1.  The resulting curve has the form \begin{equation}
Q_{2n}(z) + y Q_{2n-1}(z) + \cdots + y^{2n-1} Q_1(z) +  y^{2n}=0 \label{B}
\end{equation}
in terms of the invariant coordinates $x=s^2$, $y=v^2$ and $z=sv$, with
the differential given by $\lambda=zdy/y$. 
Here, $Q_{j}(z)$ is a polynomial of degree (at most) $j$.
From our construction, we see that the coefficient of $z^j$ of $Q_j(z)$ encodes a coupling. This coefficient is zero when $j$ is odd. The coefficients of $Q_{2n}(z)$ is the mass parameter for $\SU(2n)$  flavor symmetry, and  the coefficients of other $Q_j(z)$ are Coulomb branch parameters. 
This is the standard Seiberg-Witten curve of a quiver gauge theory with gauge group $\U(2n-1)\times \U(2n-2)\times \cdots \times \U(1)$, with bifundamental hypermultiplets between two consecutive gauge groups, and $2n$ additional flavors for the $\U(2n-1)$ group, with a special choice of the coupling constants.  
Let us call this the matter sector B.
\item In the region $  \epsilon^{1/2} \ll |s|  \ll 1 $,  only the coefficients $u_{2n-l,2k-l}$ survive. The curve is simply \begin{equation}
z^{2n} + u_{2n-1,2k-1} z^{2n-1} + \cdots + u_{0,2k-2n} =0
\end{equation} with the differential $\lambda=zds/s$.  This tube generates an $\SU(2n)$ gauge multiplet, connecting the $\SU(2n)$ flavor symmetries of the two sectors A, B given above.  
\end{itemize}
From this, we find that the physics at the singularity is given by an infrared free $\SU(2n)$ gauge theory coupled to the matter sector A and B.  

We can also turn on the $\SO(4)$ mass parameters in the analysis. They are the mass terms for the flavor symmetry $\SU(2)\times \SU(2)$ for the parts $k^2$  and $(k-n)^2$ of the two punctures of the matter sector A.   We can check that the $\SO(4)$ mass parameters do not modify the matter sector B.  One of the two mass parameters deform the orbifold singularity at $t=-1$, which clearly does not affect the sector B. The other mass parameter $\mu$ deform the singularity at $t=1$, and modify the relations between the variables $x$, $y$, $z$ introduced above to $xy = z^2 +\mu$. The curve \eqref{B} of the sector B is written purely in terms of $y$ and $z$, and the differential is still $\lambda=zdy/y$. Therefore the mass parameter $\mu$ does not affect the sector B either. 

By coupling $\USp(2k)$ gauge multiplet and two additional fundamental hypermultiplets to the matter sector A via the flavor symmetry $\SU(2k)$ associated to the puncture $[1^{2k}]$, we realize the theory shown in the third row of Fig.~\ref{fig:summaryn}. This is what we wanted to demonstrate. 

\section{Conclusions and speculations}\label{sec:conclusions}
In this paper, we considered $k$ D3-branes probing four D7-branes and an O7-plane on the orbifold $\bC^2/\bZ_{2n}$ and on the smooth ALE space $\widetilde{\bC^2/\bZ_{2n}}$.  For technical reasons, we chose the holonomy at infinity to be $\diag(+,+,+,+,-,-,-,-)$. 

On the orbifold, the worldvolume theory on the D3-branes is a 4d $\cN=2$ supersymmetric theory with gauge group  $\USp(2k_+)\times \prod_{i=1}^{n-1}\U(k_i)\times \USp(2k_-)$ with bifundamental hypermultiplets between two consecutive gauge groups, and two fundamental hypermultiplets for each of the two $\USp$ groups. 
 We chose in particular the case  \begin{equation}
\USp(2k)\times \U(2k)\times \cdots\times \U(2k)\times \USp(2k) \label{X}
\end{equation} which corresponds to the holonomy at the origin $\diag(+,+,+,+,-,-,-,-)$, and the case \begin{equation}
\USp(2k)\times \U(2k+2)\times \cdots\times \U(2k+2n-2)\times \USp(2k+2n) \label{Y}
\end{equation} which corresponds to the  holonomy at the origin $\diag(+,+,+,+,+,+,+,+)$. 

In the former case \eqref{X},  we analyzed the system using string duality, and found that the blow-up parameters can be introduced only in a strongly-coupled limit. There, we have  weakly-coupled dual gauge multiplets  of the form $\U(2n)\times  \U(2n-2)\times \cdots \times \U(2)$. We argued that giving non-zero blow-up parameters requires that we replace this chain of gauge multiplets with another chain, $\U(2n)\times  \U(2n-1)\times \cdots \times \U(1)$, and that the blow-up parameters are the FI parameters for these gauge multiplets.  The result is shown in the third row of Fig.~\ref{fig:summaryn}.

In the latter case \eqref{Y}, we analyzed the system field theoretically, along the line of \cite{Gaiotto:2010jf,Giacomelli:2012ea}. One of the gauge group, $\USp(2k+2n)$, is asymptotically free, and at one of two most  singular points on the Coulomb branch, we can find an infrared dual description,  which is again the theory shown in the third row of Fig.~\ref{fig:summaryn}. Here the author does not currently have a direct argument to show that the FI terms of the unitary gauge multiplets correspond to the blow-up parameters.

As a result, we have a description of the moduli space of $\SO(8)$ instantons on the smooth ALE space $\widetilde{\bC^2/\bZ_{2n}}$ with the holonomy at infinity being $\diag(+,+,+,+,-,-,-,-)$ in terms of a hyperk\"ahler quotient of a flat space times the Higgs branch of a class S theory.  As a holomorphic symplectic manifold, this construction is mathematically completely explicit, assuming the result in an unpublished work \cite{GinzburgKazhdan}. When $k$ is sufficiently small, we can give an explicit description even without assuming the content of \cite{GinzburgKazhdan}. 

Let us discuss how we might extend our analysis to larger SO groups. Our first method which  used the string dualities is not very adequate, as our argument relied  on the fact that the gauge theory is superconformal and there is no bending  of the NS5-branes.  Our second method which used the field-theoretical duality should be applicable, although we do not have a direct way to show that the FI terms in the infrared dual description are the blow-up parameters.  The field-theoretical duality employed is the one studied in \cite{Gaiotto:2010jf,Giacomelli:2012ea}, and is not currently developed sufficiently enough to allow us to analyze this general case. Hopefully this will change in the near future. 

A natural question is whether class-S technique can be used to study instanton moduli spaces  of groups other than SO groups on smooth ALE spaces.  The unitary groups might look easier than the orthogonal groups, for example.   The standard quiver gauge theories describing unitary instantons on the ALE spaces have all the FI parameters corresponding to the blow-up parameters of the ALE space; but one can still ask if the class S technique would shed new light on the system.  Unfortunately, these quiver gauge theories often have gauge nodes that are very infrared-free, and class S constructions are at present not immediately applicable here, as they are developed thus far mainly for systems that are conformal or slightly ultraviolet-free. We need to wait until the class S technique is extended to infrared free systems.

Finally, let us speculate how we might study the moduli space of  orthogonal instantons on ALE spaces of type D and E. Note that at least for $\SO(8)$ and with the holonomy at infinity $\diag(+,+,+,+,-,-,-,-)$, we found the following structure: \begin{equation}
\tilde\cM_{A_{2n-1},k,\vec\xi} = (X_{A_{n-1},k} \times  \cO_{A_{2n-1},\vec \xi}) \hkq  \SU(2n)
\end{equation} where $\tilde\cM_{A_{2n-1},k,\vec\xi}$ is the moduli space of SO instantons on the ALE space of type $A_{2n-1}$, with the blow-up parameter $\vec\xi$, $X_{A_{2n-1},k}$ is a certain fixed hyperk\"ahler manifold, and $\cO_{A_{2n-1},\vec \xi}$ is the semisimple orbit of $\SU(2n)$ with the parameter $\vec \xi$, and $\hkq $ denotes the hyperk\"ahler quotient construction. 
The  dimension of $\cO_{A_{2n-1},\vec\xi}$ for generic $\xi$ is $(\dim \SU(2n) - \rank \SU(2n))/2$. Therefore, $X_{A_{n-1},k}$ has $(\dim \SU(2n) + \rank \SU(2n))/2$ more quaternionic dimensions than $\tilde\cM_{A_{2n-1},k,\vec\xi} $.  

When $\vec\xi$ is set to 0, $\lim_{\vec\xi\to 0}\cO_{A_{2n-1},\vec \xi} = \cN_{A_{2n-1}} $ is the nilpotent orbit of $\SU(2n)$, and we have \begin{equation}
(X_{A_{2n-1},k} \times  \cN_{A_{2n-1}}) \hkq  \SU(2n)  = \cM_{A_{2n-1},k}= V_{A_{2n-1},k} \hkq  G_{A_{2n-1},k}
\end{equation}  where $\cM_{A_{2n-1},k}$ is the moduli space of instantons on the orbifold $\bC^2/\bZ_{2n}$ with a trivial holonomy at the origin, and $(V_{A_{2n-1},k},G_{A_{2n-1},k})$ is a known pair of a vector space and a group realizing this moduli space as a hyperk\"ahler quotient of a flat space. 

It is noticeable that the objects involved, namely $\cO_{A_{2n-1},\vec \xi}$,
$\cN_{A_{2n-1}}$, $\SU(2n)$ are all naturally  associated  to the type of the ALE space.  So, for other ALE spaces of type $\Gamma=D_n$ and $E_n$, the author would speculate that  we might have the same structure, where the semisimple orbits and the nilpotent orbits involved are replaced with those of the Lie algebra of type $\Gamma$:
\begin{equation}
\tilde\cM_{\Gamma,k,\vec\xi} = (X_{\Gamma,k} \times  \cO_{\Gamma,\vec \xi}) \hkq  \Gamma, \qquad
\cM_{\Gamma,k} = (X_{\Gamma,k} \times  \cN_{\Gamma}) \hkq  \Gamma.
\end{equation} 
  In particular,  $X_{\Gamma,k}$ would have $(\dim \Gamma + \rank \Gamma)/2$ more quarternionic dimensions than $\tilde\cM_{\Gamma,k,\vec \xi}$.

%% wrote up to here during the daytime on Jan. 20. 

\section*{Acknowledgements}
The author thanks  A. Dymarsky and J. Heckman for old discussions in 2011, and D. Gaiotto for reminding me of the reference \cite{Intriligator:1997kq}, which rekindled his interest on this subject. He also would like to thank N. Mekareeya for the discussion on \cite{Dey:2013fea}, and H. Nakajima for helpful explanations on the moduli space of orthogonal instantons on the ALE spaces.  S. Cherkis and S. Giacomelli kindly read the prepreprint carefully, pointed out many typos, and suggested various improvements of the draft.  It is a great pleasure for the author to thank them.
The author is  supported in part by JSPS Grant-in-Aid for Scientific Research No. 25870159,
and in part by WPI Initiative, MEXT, Japan at IPMU, the University of Tokyo.

\bibliographystyle{ytphys}
\small\baselineskip=.9\baselineskip
\let\bbb\bibitem\def\bibitem{\itemsep1pt\bbb}
\bibliography{ref}
\end{document}